# A generalized mechanical model using stress-strain duality at large strain for amorphous polymers

C.A. Bernard [1,2,3*], D. George [3], S. Ahzi [3**], Y. Rémond [3]

[1] Frontier Research Institute for Interdisciplinary Sciences (FRIS), Tohoku University, Sendai, Japan
[2] ELyTMaX UMI 3757, CNRS–Université de Lyon–Tohoku University, International Joint Unit, Tohoku University, Sendai, Japan
[3] University of Strasbourg, CNRS, ICube Laboratory, 2 Rue Boussingault, 67000 Strasbourg, France

**Abstract**

Numerous models have been developed in the literature to simulate the thermomechanical behavior of amorphous polymer at large strain. These models generally show a good agreement with experimental results when the material is submitted to uniaxial loadings (tension or compression) or in case of shear loadings. However, this agreement is highly degraded when they are used in the case of combined load cases. A generalization of these models to more complex loads is scarce. In particular, models that are identified in tension or compression often overestimate the response in shear. One difficulty lies in the fact that 3D models must aggregate different physical modeling, described with different kinematics. This requires the use of transport operators complex to manipulate. In this paper, we propose a mechanical model for large strains, generalized in 3D, and precisely introducing the adequate transport operators in order to obtain an exact kinematic. The stress strain duality is validated in the writing of the power of internal forces. This generalized model is applied in the case of a polycarbonate amorphous polymers. The simulation results in tension/compression and shear are compared with the classical modeling and experimental results from the literature. The results highly improve the numerical predictions of the mechanical response of amorphous polymers submitted to any load case.

∗ **Corresponding author:** Dr. Chrystelle Bernard (chrystelle.bernard@rift.mech.tohoku.ac.jp)

## 1. Introduction

Amorphous polymers are known to be highly strain rate and temperature dependent. To understand their deformation mechanisms and to model their thermomechanical behavior at small and large strains, numerous studies [1–21] have been performed since the early 30's. Most of the theories developed in the literature try to understand and to model the physics and the mechanics of amorphous polymers leading to the deformation processes. Thus, they have highlighted the importance of the different relaxation phenomena (glass transition and secondary relaxations) on the mechanical behavior of these materials, in particular the strain rate and temperature dependencies of the elastic modulus [1–3] and the yield stress [6–8, 11, 15]. Other models focus on the orientational hardening observed at large strain in the mechanical behavior of polymers [16–19, 22–24]. Based on these different theories, several numerical (visco)elastic-viscoplastic models [25–29] have been developed using a decomposition of the total deformation gradient into an elastic and an inelastic parts. At large strain, these models exhibit a good correlation between experimental results and numerical predictions for simple load cases such as uniaxial loadings (compression or tension), or independently in shear loading. The main

---

** Current address: Qatar Environment and Energy Research Institute, Hamad Bin Khalifa University, Qatar Foundation, PO Box 5825, Doha, Qatar

reason for this is because the model parameters are identified on these specific loading cases (uniaxial compressive/tensile loadings or shear loadings). However, when the same set of parameters is used to predict both uniaxial and shear mechanical behavior, inaccuracy of the numerical mechanical response is observed. Since large strain involves internally complex movements and deformation mechanisms, a true kinematics of the deformable solid is needed to predict the numerical mechanical behavior of amorphous polymers under any loading cases. However, introducing the true kinematics of the deformable solid into the constitutive equations is not an easy task. It needs to take into account transport operators such as rigid body spins induced by the different strain measures used. The solution proposed in this study is not unique, nevertheless, it allows to fit experimental results while focusing on a clear and strict theoretical background for the modelling of the large strain behavior of amorphous polymers.

Under the assumption of small strain, there exist a linearized deformation tensor $\varepsilon$ such that $\dot{\varepsilon} = \mathbb{D}$, where $\dot{\varepsilon}$ is the total time derivative of the linearized deformation tensor $\varepsilon$ and $\mathbb{D}$ is the rate of deformation tensor, defined as the symmetric part of the velocity gradient $\mathbb{L}$. However, at large strain, such simple relation cannot be found due to the difficulty to calculate the material time derivative of the strain measure in the general three-dimensional deformation case. Among the various strain measures available in the literature, the logarithmic strain introduced by Hencky [30] was historically favored as it is considered as the most adequate strain measure in particular for experimental applications and due to its additive property [31–33]. Nevertheless, in the presence of fixed principal axes of strain, a rather simple expression links the rate of the deformation tensor $\mathbb{D}$ and the logarithmic strain of the left stretch tensor $\mathbb{V}$:

$$\mathbb{D} = \overline{(\ln \mathbb{V})^{\cdot}}, \qquad (1)$$

with $\overline{(\ )^{\cdot}}$ the time derivative of a given tensor. However, in the general case, when the principal axes of strain are not fixed, this relationship is no longer valid. Several authors [34–46] have theoretically investigated how to link $\mathbb{D}$ to the logarithmic strain in the general three dimensional strain. By considering the rotation of the Eulerian or Lagrangian ellipsoid, Hill [36] established a relationship between $\mathbb{D}$ and the logarithm of the right stretch tensor $\mathbb{U}$ when the principal stretches are distinct. Later, Gurtin and Spear [37] proposed a relationship between $\mathbb{D}$ and $\ln \mathbb{V}$:

$$\mathbb{D} = (\ln \mathbb{V})^{\circ} - sym(\mathbb{F}\Omega_r \mathbb{F}^{-1}) \qquad (2)$$

where $\mathbb{F}$ is the deformation gradient, $\Omega_r$ is the spin of the right principal axes of strain. $(\ln \mathbb{V})^{\circ}$ is the time derivative of $\ln \mathbb{V}$ given by the corotational derivative or Jaumann rate formula, and $sym(\mathbb{F}\Omega_r \mathbb{F}^{-1})$ corresponds to the symmetric part of $\mathbb{F}\Omega_r \mathbb{F}^{-1}$. Hoger [38, 39] proposed a rather complicated formula to relate $\mathbb{D}$ and the derivative of $\ln \mathbb{V}$ and $\ln \mathbb{U}$ A good correlation is obtained to Hill's results. Moreover, Hoger [38, 39] established a necessary and sufficient condition for the Jaumann and the corotational derivatives to be equal to the rate of deformation tensor.

Deriving a formula for the rate of deformation tensor from the strain measure will induce modifications in the expression of the stress tensor [46–49] according to stress-strain duality. By stress-strain duality, we understand that the power of internal forces $\mathcal{P}_i$ is an intrinsic concept, independent of the chosen basis. In a given deformation basis,

$$\mathcal{P}_i = \int (-\mathbb{T} : \mathbb{D}) dv \qquad (3)$$

where $\mathbb{T}$ is the stress tensor. As a consequence of the stress-strain duality, the link between the Cauchy stress tensor and the rate of deformation tensor needs to account for the spins of the ellipsoid strain in the tensors' expressions.

In this paper, to propose a generalized model, we investigate the stress-strain duality by considering the true (3D) kinematics of the deformable solid and its consequences on the numerical thermomechanical modelling of amorphous polymers at large strain. Thus, we implement one of the theories developed to correlate the rate of deformation tensor to the time derivative of the logarithmic strain into an elastic-viscoplastic model from the literature. We choose the polymer physics based model developed by Richeton et al. [26] which allows to reproduce with accuracy, for uniaxial loading cases, the thermomechanical behavior of amorphous polymers over a wide range of strain rates and temperatures. This model is briefly detailed in section 2. Section 3 is dedicated to the introduction of the true (3D) kinematics of the deformable solid according to the formalism of the numerical model. Then, a comparison between experimental results from the literature and the numerical predictions is provided in section 4. We show that by integrating the true (3D) kinematics of the deformable solid into the constitutive equations of amorphous polymers model, we develop a generalized model able to predict both uniaxial and shear loading using the same set of parameters.

## 2. Constitutive equations

Before presenting the generalized model in Section 3, we recall the previous work of Richeton's model [26]. Richeton's model [26] is an elastic-viscoplastic model based on the time–temperature superposition principle of the (i) elastic modulus [3] and (ii) yield stress [15]. To account for the drop of stiffness due to the glass transition, the expression of the elastic modulus considers Weibull statistics to represent its evolution over a large range of temperatures from glassy to rubbery region [3]. The evolution of the yield stress as a function of the temperature and the strain rate is described by the cooperative model which assumes cooperative movements between the molecular chain segments allowing them to jump from one equilibrium position to another. Based on previous work from Fotheringham and Cherry [13, 14], Richeton [15] assumed (i) the existence of an internal stress depicting the thermal history of the material and (ii) that the flow of the polymer is allowed by the cooperative movement of several polymer chain segments.

These two models were implemented into a general 3D framework based on (i) the Cauchy stress tensor governed by the logarithmic strain measure to model the elastic behavior, and (ii) the back-stress tensor governed by the Green-Lagrange strain measure to model the viscoplastic behavior of amorphous polymers. In this section 2, the main equations of the constitutive model developed by Richeton [26] are presented. They will be used to introduce the stress-strain duality in section 3.

### 2.1. Preliminary kinematics

Let us consider a deformable solid where each material point will move from one position to another during its deformation through various configurations (initial, intermediate and final). At initial time $t_0$, let $\boldsymbol{X} \in \mathbb{R}^3$ be the spatial position of a material point, in the initial (or reference) configuration $\mathcal{R}_{t_0}^0 \subset \mathbb{R}^3$, contained in the deformable solid. At time $t$, the particle is in the final (or current) configuration $\mathcal{R}_t \subset \mathbb{R}^3$. Let us denote $\boldsymbol{x} \in \mathbb{R}^3$ the spatial position of the material point in this final configuration. To follow the deformation of the material point from $\mathcal{R}_{t_0}^0$ to $\mathcal{R}_t$ over the time interval $\Delta t = t - t_0$, the deformation gradient $\mathbb{F} = \nabla x(\boldsymbol{X}, t)$ is used. This total deformation can be seen as a combination of elastic 'e', thermal 'th' and plastic or viscoplastic 'p' deformations, each one of these parts linking one configuration (initial or intermediate) to the next configuration (intermediate or final). The different intermediate and final configurations are related to the spatial position induced by a specific deformation. Let us assume the existence of two intermediate configurations denoted $\mathcal{R}_t^1$ and $\mathcal{R}_t^2$. $\mathbb{F}^p$ represents the plastic deformation gradient related to the plastic or viscoplastic deformation of the solid particle, and links together the initial and the first intermediate configurations, $\mathcal{R}_{t_0}^0$ and $\mathcal{R}_t^1$ respectively. Similarly, $\mathbb{F}^{th}$ is the thermal deformation gradient, defined from the first intermediate configuration $\mathcal{R}_t^1$ to the second $\mathcal{R}_t^2$. It allows describing the evolution of the thermal deformation over time. $\mathbb{F}^e$ is the elastic deformation gradient, defined from $\mathcal{R}_t^2$ to $\mathcal{R}_t$. This last gradient allows describing the evolution of the elastic deformation over time. The different gradient tensors are obtained by multiplicatively decomposing the total deformation gradient $\mathbb{F}$ such that $\mathbb{F} = \mathbb{F}^e \mathbb{F}^{th} \mathbb{F}^p$. A schematic representation of the different configurations and deformation tensors is represented in Figure 1.

The solid deformation can occur through stretches and/or rotations. To identify the influence of each contribution, the polar decomposition, which multiplicatively decomposes the left or right stretch tensors, respectively $\mathbb{U}$ or $\mathbb{V}$, from the rotation tensor $\mathbb{R}$, is performed on the deformation gradient. Thus, the deformation gradient is written as $\mathbb{F} = \mathbb{R}\mathbb{U} = \mathbb{V}\mathbb{R}$. The rotation tensor $\mathbb{R}$ links together the right and the left stretch tensors $\mathbb{U}$ and $\mathbb{V}$, respectively associated with the initial basis of deformation $\boldsymbol{\Delta}$, and the final basis of deformation $\boldsymbol{\delta}$ through $\mathbb{U} = \mathbb{R}^T \mathbb{V} \mathbb{R}$ or $\mathbb{V} = \mathbb{R} \mathbb{U} \mathbb{R}^T$. The principal stretches are the same whatever the basis. Thus, $\mathbb{U}$ and $\mathbb{V}$ have the same eigenvalues. However, due to the change of configuration between $\mathbb{U}$ and $\mathbb{V}$, their orthogonal basis of eigenvectors, respectively $\boldsymbol{\Delta}$ and $\boldsymbol{\delta}$, are different. They are linked together through the rotation tensor $\mathbb{R}$ via the relation $\boldsymbol{\delta} = \mathbb{R}\boldsymbol{\Delta}$.

In addition, the current spatial position of the material point is correlated to its past deformation history. This is of a high importance for polymers since it can strongly affect the microstructure and the thermomechanical behavior of the material. The past and the current spatial position of the material point are related through the velocity gradient $\mathbb{L} = \dot{\mathbb{F}}\mathbb{F}^{-1}$, where the superposed dot ( $\dot{}$ ) designates the material time derivative of a given tensor and $\mathbb{F}^{-1}$ represents the inverse tensor of $\mathbb{F}$.

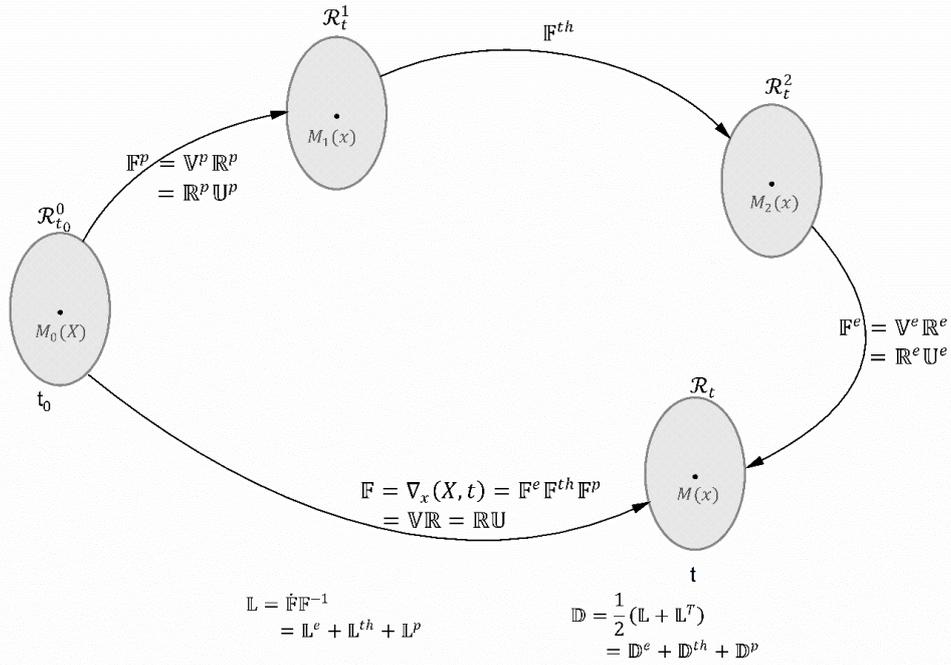

Figure 1: Schematic representation of the different configurations, deformation tensors and deformation gradients. $\mathcal{R}_{t_0}^0$ is the initial (or reference) configuration, $\mathcal{R}_t^1$ and $\mathcal{R}_t^2$, the first and second intermediate configurations, respectively and $\mathcal{R}_t$ is the final (or current) configuration. The rate of deformation tensor $\mathbb{D} = \frac{1}{2}(\mathbb{L} + \mathbb{L}^T) = \mathbb{D}^e + \mathbb{D}^{th} + \mathbb{D}^p$, is defined as the symmetric part of the velocity gradient $\mathbb{L} = \dot{\mathbb{F}}\mathbb{F}^{-1} = \mathbb{L}^e + \mathbb{L}^{th} + \mathbb{L}^p$. Using the polar decomposition, the deformation gradient $\mathbb{F}$ is multiplicatively decomposed into rotation tensor $\mathbb{R}$ and left or right stretch tensors, respectively $\mathbb{U}$ or $\mathbb{V}$.

The behavior of amorphous polymers being strongly strain rate dependent, the rate of the stretches, the material properties and model parameters need to be evaluated to compute the flow rule of the material. From the expression of $\mathbb{L}$, a unique additive decomposition can be performed such that $\mathbb{L} = \mathbb{D} + \mathbb{W}$ where $\mathbb{D}$, the symmetric part of $\mathbb{L}$, is related to the rate of the deformation tensor and $\mathbb{W}$, the skew-symmetric part of $\mathbb{L}$, is the spin tensor. $\mathbb{W}$ is related to the rate of the rotation or vorticity of the solid. In order to identify the contributions of each configuration to the rate of stretches, the elastic, thermal and plastic deformation gradients are introduced in the expression of $\mathbb{D}$. Thus, the rate of deformation tensor $\mathbb{D}$ is additively decomposed into elastic, thermal and plastic rate of deformation tensors, respectively $\mathbb{D}^e$, $\mathbb{D}^{th}$ and $\mathbb{D}^p$:

$$\mathbb{D} = \mathbb{D}^e + \mathbb{D}^{th} + \mathbb{D}^p \tag{4}$$

where $\mathbb{D}^e$, $\mathbb{D}^{th}$ and $\mathbb{D}^p$ are respectively the symmetric part of the elastic part $\mathbb{L}^e$, the thermal part $\mathbb{L}^{th}$, and the plastic part $\mathbb{L}^p$ of the velocity gradient $\mathbb{L} = \mathbb{L}^e + \mathbb{L}^{th} + \mathbb{L}^p$ with

$$\begin{cases} \mathbb{L}^e &= \dot{\mathbb{F}}^e \mathbb{F}^{e-1} \\ \mathbb{L}^{th} &= \mathbb{F}^e \dot{\mathbb{F}}^{th} \mathbb{F}^{th-1} \mathbb{F}^{e-1} \\ \mathbb{L}^p &= \mathbb{F}^e \mathbb{F}^{th} \dot{\mathbb{F}}^p \mathbb{F}^{p-1} \mathbb{F}^{th-1} \mathbb{F}^{e-1} \end{cases} \quad (5)$$

In equation (5), $\mathbb{F}^e$, $\mathbb{F}^{th}$, and $\mathbb{F}^p$ are defined in the schematic decomposition of Figure 1. By decomposing the rate of deformation tensor into its elastic, thermal and plastic parts, we can investigate the rigid body spins due to each deformation. Thus, the influence of each strain measure can be taken into account when modelling the thermomechanical behavior of the material.

### 2.2. Elastic-viscoplastic model

In Richeton's model [26], according to Richeton et al. [3], Dupaix and Boyce [50], Pandini and Pegoretti [51, 52] and Bernard et al. [53], the Young's modulus and Poisson's ratio are strongly strain rate and temperature dependent. Therefore, the fourth order rigidity tensor $\mathbb{C}^e$ is also strain rate and temperature dependent through these two parameters [3]. The elastic response of the material is then described by the Cauchy stress tensor $\mathbb{T}$, defined using the logarithm of the strain measure $\mathbb{V}^e$ with $\mathbb{F}^e = \mathbb{V}^e \mathbb{R}^e$:

$$\mathbb{T} = \frac{1}{J} \mathbb{C}^e \ln \mathbb{V}^e \quad (6)$$

where $J = \det \mathbb{F}^e$ is the volume change. The Cauchy stress tensor is computed in the orthogonal basis of deformation associated with $\mathbb{V}^e$. Thus, $\mathbb{T}$ is defined in the current configuration $\mathcal{R}_t$ since it corresponds to the final orthogonal basis of the elastic deformation.

Beyond the yield point, predicted by the cooperative model [15], the mechanical behavior of the material is no longer elastic. According to Bowden and Haward [54], Boyce et al. [55], the stress-strain response of glassy polymers and the associated physics, particularly the orientational hardening, are close to the behavior observed for rubbery materials like elastomers. Thus, to account for the hardening observed in the viscoplastic response, the glassy polymers can be considered as a rubbery network. To represent this behavior, several hyperelastic models have been developed by considering a statistical model for the distribution of the polymer chain segments in the rubbery network [16–19, 22, 56]. These models allow predicting the sharp increase of stress due to the chain length locking. Among them, the 8-chain model developed by Arruda and Boyce [18] provides an accurate description of this mechanical behavior. When the material is submitted to a displacement or stress field, the polymer chain segments will move according to the imposed deformation until they break or they reach their maximal extension. To describe the orientational hardening of the material, the 8-chain model considers only two parameters: $C_R$, the rubbery modulus, and $N$, the number of statistical links in the chain (which is linked to the locking stretch). The viscoplastic response of the material is described using a kinematic hardening approach where the back-stress tensor $\mathbb{B}$ is defined using the 8-chain model:

$$\mathbb{B}_i = \frac{C_R}{3\lambda_{chain}} \sqrt{N} \mathcal{L}^{-1}\left(\frac{\sqrt{N}}{\lambda_{chain}}\right)(\lambda_i^2 - \lambda_{chain}^2) \quad \text{with} \quad \mathbb{B} = \boldsymbol{\delta}^p [\![\mathbb{B}_i]\!] \boldsymbol{\delta}^{p-1}. \quad (7)$$

Here, $\mathbb{B}_i$ are the principal components of the back-stress. Therefore, $\mathbb{B}_i$ describes the diagonal components of the back-stress tensor $\mathbb{B}$ obtained after multiplication by the transfer matrix $\boldsymbol{\delta}^p$. $\lambda_i$ are the principal plastic stretches (eigenvalues of $\mathbb{V}^p$ with $\mathbb{F}^p = \mathbb{V}^p \mathbb{R}^p$), and $\lambda_{chain}$ is the chain stretch equal to $(\lambda_1^2 + \lambda_2^2 + \lambda_3^2)/3$. The notation $[\![\ ]\!]$ defines the given tensor in the principal axes. The back-stress tensor is therefore computed in the orthogonal basis of deformation associated with $\mathbb{V}^p$. The strain measure used in the back-stress relationship is relative to the Green-Lagrange strain measure $\mathbb{E}$. More details about the actual strain measure will be provided in section 3.2. Thus, $\mathbb{B}$ is defined in the first intermediate configuration $\mathcal{R}_t^1$ since it corresponds to the final orthogonal basis of the viscoplastic deformation.

To determine the rate of deformation tensors, we introduce the driving stress tensor $\mathbb{T}^*$ defined as the difference between the Cauchy stress and the back-stress. Richeton et al. [26] proposed to rotate $\mathbb{T}$ in order to define the Cauchy stress tensor and the back-stress tensor in the same first intermediate configuration $\mathcal{R}_t^1$:

$$\mathbb{T}^* = \mathbb{R}^{eT}\mathbb{T}\mathbb{R}^e - \mathbb{B} \tag{8}$$

where $\mathbb{R}^e$ is the elastic rotation tensor.

The flow rule allows expressing the plastic rate of deformation tensor $\mathbb{D}^p$ as follows:

$$\mathbb{D}^p = \frac{\dot{\gamma}^p}{\sqrt{2}\tau} dev(\mathbb{T}^*) \tag{9}$$

with $\dot{\gamma}^p$ the plastic shear strain rate computed from the expression of the cooperative model given by Richeton et al. [15, 57], $\tau$ is the effective equivalent stress obtained from the deviatoric part of the driving stress tensor $dev(\mathbb{T}^*)$ ($\tau = [dev(\mathbb{T}^*):dev(\mathbb{T}^*)/2]^{1/2}$).

Finally, the thermal deformations occur through conduction, convection or self-heating phenomena. The evolution of the thermal rate of deformation tensor is assumed to follow an isotropic thermal expansion. Thus, the tensor $\mathbb{D}^{th}$ associated with the thermal deformation is given by:

$$\mathbb{D}^{th} = \beta(\theta)\dot{\theta}\mathbb{I} \tag{10}$$

where $\beta$ is the thermal expansion coefficient, dependent on the temperature $\theta$, and $\mathbb{I}$ is the identity tensor.

The application of this model is conducted on polycarbonate (PC) and the model parameters (materials properties and parameters) are defined in the work of Richeton et al. [26] and are given in Table 1.

**Table 1: Material properties for polycarbonate (PC) from Richeton et al. [26]**

|  |  | PC |
|---|---|---|
| Elastic properties | $\dot{\varepsilon}^{ref}$ (s$^{-1}$) | 1 |
|  | $E_1^{ref}/E_2^{ref}/E_3^{ref}$ (MPa) | 3500/1700/20 |
|  | $T_\beta^{ref}/T_g^{ref}/T_m^{ref}$ (K) | 195/423/436 |
|  | $m_1/m_2/m_3$ | 5/80/15 |
|  | s | 0.011 |
| Flow properties | n | 5.88 |
|  | V (m$^3$) | $9.18 \times 10^{-29}$ |
|  | $\tau_i(0)$ (MPa) | 81 |
|  | m (MPa K$^{-1}$) | 0.14 |
|  | $\dot{\gamma}_0$ (s$^{-1}$) | $1.36 \times 10^{13}$ |
|  | $\Delta H_\beta$ (kJ mol$^{-1}$) | 40 |
|  | $c_1$ | 16.19 |
|  | $c_2$ (K) | 55.6 |
|  | $\tau_{ps}/\tau_i$ | 0.57 |
|  | h (MPa K$^{-1}$) | 300 |
| 8-chain model properties | $C_R(0)$ (MPa) | 37.57 |
|  | a (MPa K$^{-1}$) | 0.077 |
|  | $N(0)$ | 1.960 |
|  | b | 0.0013 |
| Material properties | $\rho$(298 K) (kg m$^{-3}$) | 1200 |
|  | $\Gamma$(298 K) (W m$^{-1}$ K$^{-1}$) | 0.187 |
|  | $c_p$(298 K) (J kg$^{-1}$ K$^{-1}$) | 1200 |
|  | $\beta$(298 K) (K$^{-1}$) | $70.2 \times 10^{-6}$ |

| | $\nu(298\ K)$ | 0.36 |

Up to this point, the general background of the thermomechanical modelling of amorphous polymers behavior [3, 15] was presented. In the following, we focus on the numerical modelling of the amorphous polymer mechanical behavior without thermal activation. The development of the thermal sensitivity of our model, in the framework of the generalized modelling, will be pursued in a future work.

## 3. Generalized model

In the model presented in section 2, two different strain measures are used in eq. (6) and in eq. (7). Eq. (6) defines Cauchy stress via linear Hooke's law and using the logarithmic elastic strain (Henky strain). Eq. (7) defines the back-stress for kinematic hardening during plastic deformation via a non-linear, rubber-like, behavior where the Green-Lagrange measure of strain is used. Both of these strain measures, used to represent the large deformation of the material, will play an important role on the generalization of the numerical mechanical model developed by Richeton et al. [26]. In the following, the link between $\mathbb{D}$, the time derivative of the strain measures, and the rigid body spin tensor will be done according to these measures. In order to simplify the equations, all relationships will be written based on the right stretch tensor $\mathbb{U}$ ($\mathbb{U}^e$ and $\mathbb{U}^p$).

### 3.1. Logarithmic strain measure

Let us consider the influence of the stress-strain duality on the elastic deformation modelling in this generalized model. According to eq. (6), the Cauchy stress tensor is defined using the logarithmic strain measure of the left stretch tensor $\mathbb{V}^e$ which is suitable for large strains [47]. Several theories [47, 36–38, 44, 45] have been developed to correlate the rate of the logarithmic strain measure to the rate of deformation tensor at large strain. These theories are generally written using the right stretch tensor $\mathbb{U}$. One of these first theories has been developed by Hill [36] where he established a relationship between $\mathbb{D}$ and $\overline{\ln \mathbb{U}}$.

For applications to amorphous polymers at large strain, we propose to develop thereafter the reasoning in accordance with the logarithm strain measure.

Let $L^e = \ln \mathbb{U}^e$ where the eigenvalues of $\mathbb{U}^e$ are denoted by $\lambda_i^e$ (i=1,2,3) and the orthonormal basis of eigenvectors of $\mathbb{U}^e$ is denoted by $\mathbf{\Delta}^e$. Let $\Omega^{\mathbf{\Delta}^e}$ be the spin tensor associated to $\mathbb{U}^e$. $\mathbb{U}^e, L^e$ and $\Omega^{\mathbf{\Delta}^e}$ are defined in [47] by:

$$\begin{cases} \mathbb{U}^e &= \mathbf{\Delta}^e [\![\lambda_i^e]\!] \mathbf{\Delta}^{e-1} \\ L^e &= \mathbf{\Delta}^e [\![\ln \lambda_i^e]\!] \mathbf{\Delta}^{e-1} \\ \Omega^{\mathbf{\Delta}^e} &= -\Omega^{\mathbf{\Delta}^e} = \dot{\mathbf{\Delta}}^e \mathbf{\Delta}^{e-1} \end{cases}. \tag{11}$$

The notation $[\![\ ]\!]$ defines a diagonal tensor. Thus, $[\![\lambda_i^e]\!]$ represents the tensor $\mathbb{U}^e$ in its principal axes of deformation. Thus, the time derivative of $\mathbb{U}^e$ is given by:

$$\dot{\mathbb{U}}^e = \mathbf{\Delta}^e [\![\dot\lambda_i^e]\!] \mathbf{\Delta}^{e-1} + \Omega^{\mathbf{\Delta}^e} \mathbb{U}^e - \mathbb{U}^e \Omega^{\mathbf{\Delta}^e}. \tag{12}$$

Thus, for the time derivative of $L^e$, we obtain:

$$\dot{L}^e = \mathbf{\Delta}^e [\![\dot\lambda_{\underline{i}}^e \lambda_{\underline{i}}^{e-1}]\!] \mathbf{\Delta}^{e-1} + \Omega^{\mathbf{\Delta}^e} \ln \mathbb{U}^e - \ln \mathbb{U}^e \Omega^{\mathbf{\Delta}^e}. \tag{13}$$

The underlined subscripts are used for non-summation on the repeated indices. In terms of components in the basis $\mathbf{\Delta}^e$, the derivative of the logarithmic strain measure $\dot{L}^e$ is expressed as:

$$\dot{L}_{ij}^e = \begin{cases} \dot\lambda_i^e \lambda_j^{e-1} & \text{if } i = j \\ \Omega_{ij}^{\mathbf{\Delta}^e} \ln\left(\dfrac{\lambda_j^e}{\lambda_{\underline{i}}^e}\right) & \text{if } i \neq j \end{cases} \tag{14}$$

For establishing a relationship between the time derivative of the strain measure $\dot{\mathbb{L}}^e$ and the rate of elastic deformation tensor $\mathbb{D}^e$, we have to express $\mathbb{D}^e$ in terms of stretch and rotation tensors. $\mathbb{D}^e$ corresponds to the symmetric part of the velocity gradient $\mathbb{L}^e = \dot{\mathbb{F}}^e \mathbb{F}^{e-1}$. Using the polar decomposition of the deformation gradient $\mathbb{F}^e$, $\mathbb{D}^e$ becomes:

$$\mathbb{D}^e = \frac{1}{2}\left(\mathbb{R}^e \dot{\mathbb{U}}^e \mathbb{U}^{e-1} \mathbb{R}^{eT} + \mathbb{R}^e \mathbb{U}^{e-1} \dot{\mathbb{U}}^e \mathbb{R}^{eT}\right) \tag{15}$$

with $\mathbb{R}^e$ the elastic rotation tensor. Multiplying expression (15) by $\mathbb{R}^{eT}$ on the left side and $\mathbb{R}^e$ on the right side, and substituting $\dot{\mathbb{U}}^e$ by its expression given by eq. (12), we obtain:

$$\mathbb{R}^{eT} \mathbb{D}^e \mathbb{R}^e = \boldsymbol{\Delta}^e \left[\!\left[\dot{\lambda}_{\underline{i}}^e \lambda_{\underline{i}}^{e-1}\right]\!\right] \boldsymbol{\Delta}^{e-1} + \frac{1}{2}\left(\mathbb{U}^{e-1} \Omega^{\Delta^e} \mathbb{U}^e - \mathbb{U}^e \Omega^{\Delta^e} \mathbb{U}^{e-1}\right) \tag{16}$$

In terms of components in the basis $\Delta$, $\mathbb{D}^e$ is expressed as:

$$\mathbb{D}^e_{ij} = \begin{cases} \dot{\lambda}_i^e \lambda_j^{e-1} & \text{if } i = j \\ \dfrac{1}{2}\Omega_{ij}^{\Delta^e}\left(\dfrac{\lambda_{\underline{j}}^e}{\lambda_{\underline{i}}^e} - \dfrac{\lambda_{\underline{i}}^e}{\lambda_{\underline{j}}^e}\right) & \text{if } i \neq j \end{cases} \tag{17}$$

Expressing $\Omega^{\Delta^e}$ as a function of $\mathbb{D}^e$ given by eq. (17), replacing it in eq. (14) and using elastic deformation formalism, we obtain:

$$\left(\overline{\ln \mathbb{U}^e}\right)_{ij} = K_{\underline{ij}}^e \mathbb{D}_{\underline{ij}}^e \quad \text{with} \quad K_{\underline{ij}}^e = \begin{cases} 1 & \text{if } i = j & (a) \\ \dfrac{2\lambda_{\underline{i}}^e \lambda_{\underline{j}}^e}{\lambda_{\underline{j}}^{e^2} - \lambda_{\underline{i}}^{e^2}} \ln\left(\dfrac{\lambda_{\underline{j}}^e}{\lambda_{\underline{i}}^e}\right) & \text{if } i \neq j \text{ and } \lambda_i^e \neq \lambda_j^e & (b) \\ 1 & \text{if } i \neq j \text{ and } \lambda_i^e = \lambda_j^e & (c) \end{cases} \tag{18}$$

where $K^e$ is a symmetric tensor representing the rigid body spins induced by the true kinematics of the elastic strain measure. $\lambda_i^e$ are the principal elastic stretches (eigenvalues of $\mathbb{U}^e$). In the uniaxial directions of strain, when $i = j$ (eq. (18)(a)), the elastic strain and the logarithmic strain rate measure are equivalent. However, in the shear directions of strain, when the two principal in-plane stretches (eigenvalues of $\mathbb{U}^e$) are distinct, the rate of deformation tensor components are corrected as shown by eq. (18)(b). This correction enables to link the time derivative of the logarithmic strain measure and the rate of deformation tensor. In addition, when two eigenvalues are equals, eq. (18)(b) is undetermined. Thus, to insure the continuity of strain when $\lambda_i^e = \lambda_j^e$ with $i \neq j$, it is common to use $K_{ij}^e = 1$ (eq. (18)(c)).

As the power density of internal forces should be the same regardless of the chosen basis of deformation, we have to use the stress-strain duality to correct the writing of the stress tensor in regards of the corresponding rate of deformation tensor. Thus, an update of $\mathbb{D}^e$ leads to an update of the stress tensor. By taking into account the true (3D) kinematics of the deformable solid, the power density of internal forces is written:

$$\begin{aligned}-\mathcal{P} &= \mathbb{T}_{ij}\mathbb{D}_{ij}^e \\ &= \left(\mathbb{T}_{\underline{kl}} K_{\underline{kl}}^{e-1}\right)_{ij} \left(K_{\underline{mn}}^e \mathbb{D}_{\underline{mn}}^e\right)_{ij} \\ &= \left(\mathbb{T}_{\underline{kl}} K_{\underline{kl}}^{e-1}\right)_{ij} \dot{L}^e_{ij}\end{aligned} \tag{19}$$

Again, the underlined subscripts are classically used for non-summation on the repeated indices. Thus, the update of the Cauchy stress tensor is given according to the stress-strain duality of the logarithmic strain measure.

$$\widehat{\mathbb{T}}_{ij} = K^{e^{-1}}_{\underline{ij}} \mathbb{T}_{ij} \tag{20}$$

with the tensor $K^e$ defined in eq. (18) and $\widehat{\mathbb{T}}$ the Cauchy stress tensor after considering the stress-strain duality due to the elastic deformation. It appears, from eq. (20), that no correction of the diagonal terms of Cauchy stresses is needed since the diagonal components of the tensor $K^e$ are equal to 1. Thus, the used elastic strain measure does not induce rigid body spins for uniaxial deformation. However, as shown by eq. (18)(b-c), in the case of shear deformation, a correction of the shear components of the Cauchy stress tensor is needed when the two in plane eigenvalues are distinct. Thus, this correction allows preserving the power density of "internal" forces concept in the elastic regime.

### 3.2. Green-Lagrange measure

In the framework of this generalized model, the effect of the true (3D) kinematics of the deformable solid need to be introduced in the definition of the viscoplastic deformation modelling to obtain the global mechanical behavior. According to eq. (7), the back-stress tensor, given by the 8-chain model, is defined using stretches $(\lambda_i^2 - \lambda_{chain}^2)$. Thus, a theory of viscoplasticity will be applied to define the rigid body spins, different from the one used for the Cauchy stress tensor. To account for the inaccuracy of the strain measure at large strain, a similar approach to the elastic behavior (see previous section) is developed for the plastic rate of deformation and the back-stress tensors.

To estimate the rate of deformation and back-stress tensors correction, we assume the existence of a transformation function $f$, defined from $\mathbb{R}^+$ to $\mathbb{R}^+$, representative of the viscoplastic strain measure. In order for $f$ to be a suitable strain measure, it is required to verify three conditions [47]:

- $f(1) = 0$, $f(\mathbb{U})$ is null when there is no deformation
- $\dot{f}(1) = 1$, in small strain $f(\mathbb{U}) \equiv \mathbb{U}$
- $f$ is increasing with the dilatation

To verify these three properties, a suitable function $f$ can be given, in the principal axes of deformation, by:

$$f(\lambda_i^p) = \frac{1}{2}\left(\lambda_i^{p^2} - \lambda_{chain}^{p^2}\right) \tag{21}$$

with $\lambda_i^p$ the eigenvalues of $\mathbb{U}^p$ and $\lambda_{chain}^p = \sqrt{\left(\lambda_1^{p^2} + \lambda_2^{p^2} + \lambda_3^{p^2}\right)/3}$. In the orthogonal axes of deformation relative to the strain measure $\mathbb{U}^p$, $f$ is given by:

$$f(\mathbb{U}^p) = \mathbf{\Delta}^p \llbracket f(\lambda_i^p) \rrbracket \mathbf{\Delta}^{p-1} \tag{22}$$

with $\mathbf{\Delta}^p$ eigenvectors basis of $\mathbb{U}^p$.

One of the well-known measures based on the stretches is the Green-Lagrange measure $\mathbb{E}$ defined as:

$$\mathbb{E} = \frac{1}{2}(\mathbb{F}^T \mathbb{F} - \mathbb{I}). \tag{23}$$

If the material is incompressible, which is the assumed case for the viscoplastic deformation of amorphous polymers, then $\text{trace}(\mathbb{E}) = 0$ and eq. (23) becomes:

$$\mathbb{E} = \frac{1}{2}\left(\mathbb{F}^T \mathbb{F} - \frac{\text{trace}(\mathbb{F}^T \mathbb{F})}{3}\mathbb{I}\right) \tag{24}$$

where $\text{trace}(\mathbb{F}^T\mathbb{F})/3$ corresponds to $\lambda_{chain}^2$ mentioned in eq. (21). Thus, under the above assumption, the transformation $f$ is equal to the Green-Lagrange strain measure. To better understand the influence of the viscoplastic strain measure on the thermomechanical modelling of the material behavior, the stress-strain duality consistency with the Green-Lagrange measure will be analyzed.

The derivative of the plastic strain rate measure is given by:

$$\overline{\dot{f(\mathbb{U}^p)}} = \mathbf{\Delta}^\mathrm{p}[\![f(\dot{\lambda}_i^p)]\!]\mathbf{\Delta}^{\mathrm{p}-1} + \Omega^{\Delta^\mathrm{p}}f(\mathbb{U}^p) - f(\mathbb{U}^p)\Omega^{\Delta^\mathrm{p}} \tag{25}$$

with $\Omega^{\Delta^\mathrm{p}} = \dot{\mathbf{\Delta}}^p\mathbf{\Delta}^{\mathrm{p}-1}$ [38, 47] being the spin tensor associated to the orthogonal basis $\mathbf{\Delta}^\mathrm{p}$. In terms of components in $\mathbf{\Delta}^\mathrm{p}$, using eqs. (22) and (25), we obtain:

$$\overline{\dot{f(\mathbb{U}^p)}}_{ij} = \begin{cases} \dot{\lambda}_{\underline{i}}^p \lambda_{\underline{i}}^p & \text{if } i = j \\ \dfrac{1}{2}\Omega_{ij}^{\Delta^\mathrm{p}}\left(\lambda_{\underline{j}}^{p2} - \lambda_{\underline{i}}^{p2}\right) & \text{if } i \neq j \end{cases}. \tag{26}$$

The development performed in eq. (15) to (17) for the rate of elastic deformation tensor $\mathbb{D}^e$ is also valid for the viscoplastic strain measure. Thus, using eq. (16) (superscript "$e$" is now "$p$") and eq. (26), the relationship between the derivative of the strain measure $\overline{\dot{f(\mathbb{U}^p)}}$ and the rate of plastic deformation tensor $\mathbb{D}^p$ can be deduced:

$$\overline{\dot{f(\mathbb{U}^p)}}_{ij} = \mathbb{D}_{ij}^p \lambda_{\underline{i}}^p \lambda_{\underline{j}}^p. \tag{27}$$

If $i = j$, $\mathbb{D}_{ii}^p = \dot{\lambda}_i^p \lambda_i^{p-1}$ (see eq. (17)) leading to $\overline{\dot{f(\mathbb{U}^p)}}_{ii} = \dot{\lambda}_i^p \lambda_i^p$ (eq. (26)). Thus, eq. (27) is sufficient to take into account the correction for diagonal and shear components. According to the power of internal forces (eq. (19)), the correction of the back-stress tensor $\widehat{\mathbb{B}}$, induced by the Green-Lagrange measure, is given, in terms of component, by:

$$\widehat{\mathbb{B}}_{ij} = \mathbb{B}_{ij}\lambda_{\underline{i}}^{p-1}\lambda_{\underline{j}}^{p-1}. \tag{28}$$

Expression (28) clearly shows that the true (3D) kinematics due to the viscoplastic deformation will affect the mechanical modelling of the material whatever the loading considered (uniaxial or shear loading).

In this section 3, we developed the constitutive equations to take into account the true (3D) kinematics in the model developed by Richeton et al. [26] to predict the mechanical behavior of amorphous polymers. From a mathematical point of view, we can see that the generalization of the model will affect both uniaxial and shear loadings. In the next section, a clear understanding of the influence of the true kinematics on the prediction of the mechanical behavior of amorphous polymers will be reached.

## 4. Generalized model prediction

Section 3 highlights the influence of the true kinematics of the deformable solid on the mathematical expressions of Richeton's model for the prediction of the mechanical behavior of amorphous polymers. It has been observed, from a mathematical viewpoint, that the generalization of the numerical model introduces significant changes in the expressions of the stress tensors and rate of deformation tensor, according to the stress-strain duality concept, for both uniaxial and shear loading cases.

In this section 4, we investigate the numerical prediction of the generalized model on a polycarbonate (PC) tested at 25°C and 1 s$^{-1}$. The material properties and model parameters can be found in Table 1. The numerical predictions from our model will be compared with the numerical predictions of the classical elastic-viscoplastic model developed by Richeton et al. [26] and experimental results from the literature.

### 4.1. Uniaxial compression

Figure 2a represents the experimental results of a uniaxial compressive test on a polycarbonate. On the same figure, the experimental results are compared with the numerical prediction obtained with the model developed by Richeton et al. [26]. Figure 2b presents the numerical predictions obtained by the generalized model with the same parameters identified by Richeton et al. [26] (black line). Using the same set of parameters, the simulation shows an important increase of stress as soon as the strain softening phase is over (20% strain). Therefore, it is necessary to re-identify the set of parameters for the generalized model. This is shown by the red line in Figure 2 that represents the numerical prediction of the generalized model after re-identification of the parameter set. This new simulation shows very good agreement with the experimental results.

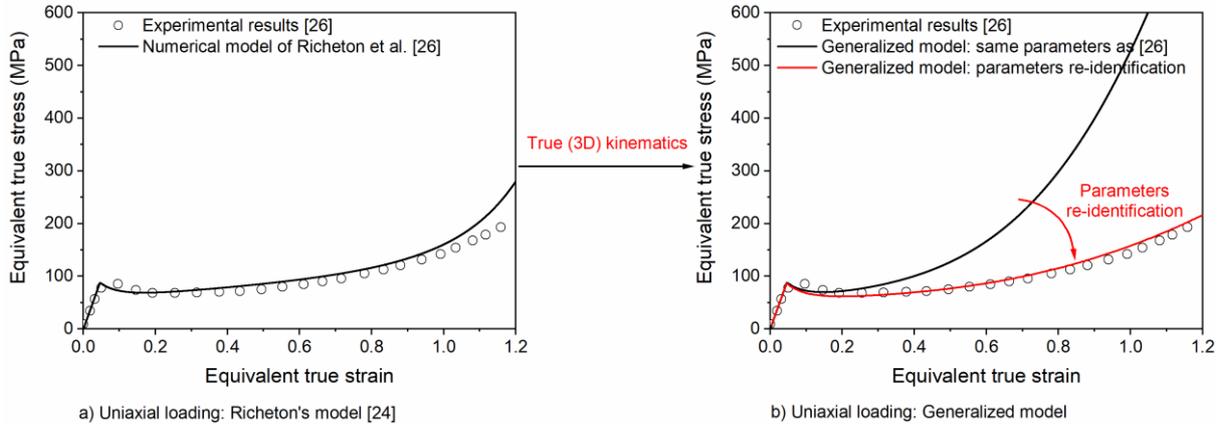

**Figure 2:** Influence of the true (3D) kinematics on the numerical prediction of the mechanical behavior of amorphous polymer submitted to uniaxial compressive loading (polycarbonate tested at 25°C and 1 s$^{-1}$). Comparison between a) the mechanical model developed by Richeton et al. [26] and b) the generalized model. Because the correction of the Green-Lagrange measure affects both diagonal and non-diagonal components, re-identification of the model parameters (see Table2) is needed to fit the experimental behavior in uniaxial compression. Both numerical predictions were compared to experimental data in uniaxial compression from Richeton et al. [26].

Since the set of parameters is generally determined to fit the experimental results in uniaxial loadings (tensile or compressive loading), it is necessary to determine a new set of parameters while accounting for the influence of the true (3D) kinematics of the deformable solid. That way, the numerical predictions can fit the uniaxial compressive experimental results. The new set of parameters is presented in Table 2.

**Table 2:** New set of parameters for the description of the thermomechanical behavior of PC while accounting for the true (3D) kinematics. Only the changed parameters are mentionned.

|  | Former value [26] | New value after accounting for the transport operators |
|---|---|---|
| $C_R(0)$ (MPa) | 37.57 | 31.57 |
| $N(0)$ | 1.96 | 11.09 |

Only the two re-identified parameters are presented in Table 2. These parameters correspond to the two parameters of the 8-chain model, related to the large strain behavior of the material: the rubbery modulus, $C_R$ and $N$, the number of statistical links of length $l$ between entanglements. According to Arruda and Boyce [18], rubber elastic materials have important capabilities to deform under stretch. Thus, they present a high value for $N$, such as 7.9 for silicon rubber, 40 for gum rubber or neoprene rubber, associated with a very low rubbery modulus (lower than 1 MPa). Although, rubber materials are usually chemically and physically crosslinked unlike glassy amorphous polymers, like polycarbonate, which exhibit only physical crosslinking. It is commonly admitted that, at large strain, the mechanical behavior of glassy amorphous polymers involves similar deformation mechanisms as rubber elastic materials. In the same paper – Arruda and Boyce [18] – , the chain locking stretch $\lambda_L$ is equal to the square root of N. Elongation at break for molded polycarbonate is estimated at 233% [58], which correspond to a chain stretch of 11.09. Using this set of parameters, a good agreement is observed between the numerical prediction and the experimental results from Richeton et al. [26].

### 4.2. Shear loading

Ghorbel and collaborators [59, 60] highlight that constitutive models are able to well describe the mechanical behavior of amorphous polymer under uniaxial compressive or tensile loadings, but largely overestimate the prediction of the shear behavior. Tomita [61] implied that this overestimation of the shear behavior is partially due to the development of internal stresses in the material leading to an evolution of the number of entangled points $N$ between uniaxial and shear loadings.

Here, it is assumed that there is no microstructural change of the material between uniaxial and shear loadings. The influence of the true (3D) kinematics on the shear behavior modelling of amorphous polymers was modelled and is reported in Figure 3. The numerical predictions were compared with the experimental results on polycarbonate from Ghorbel et al. [60] and G'Sell et al. [62]. The experimental shear stress-shear strain curves were transformed into equivalent true stress-equivalent true strain curves to be compared with the numerical predictions using the relations:

$$\begin{cases} \varepsilon_{eq} = \sqrt{3}\gamma \\ \sigma_{eq} = \tau/\sqrt{3} \end{cases} \quad (29)$$

where $\tau$ and $\gamma$ refers to the shear stress and shear strain, respectively, while $\sigma_{eq}$ and $\varepsilon_{eq}$ refers to the equivalent true stress and equivalent true strain respectively.

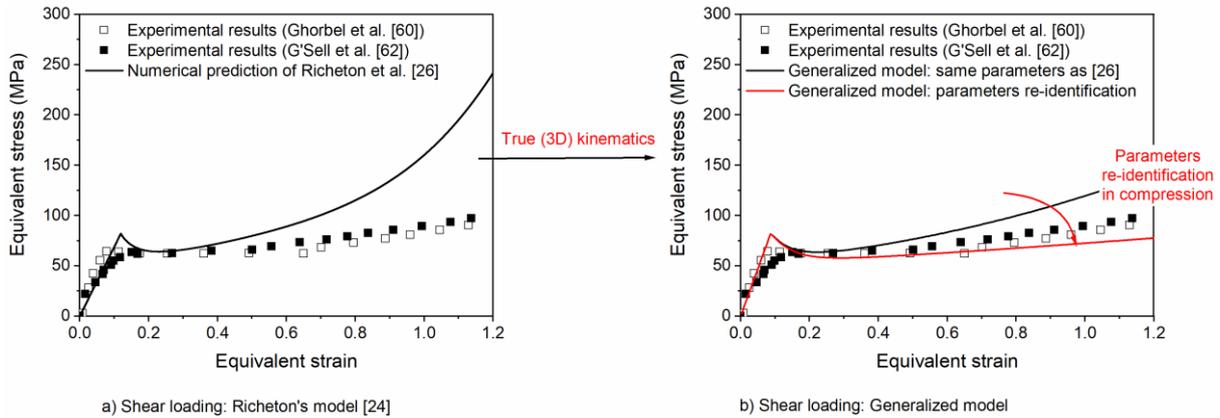

**Figure 3: Influence of the true (3D) kinematics on the numerical prediction of the mechanical behavior of amorphous polymer submitted to pure shear loading (polycarbonate tested at 25°C and 1 s$^{-1}$). Comparison between a) the mechanical model developed by Richeton et al. [26] and b) the generalized model. Accounting for the true (3D) kinematics and the parameters re-identification (see Table 2) highly improve the numerical prediction for shear loading. Both numerical predictions were compared to experimental data in shear loading from Ghorbel et al. [60] and G'Sell et al. [62].**

In Figure 3a, we used the model developed by Richeton et al. [26], identified in uniaxial loading, to simulate the mechanical behavior of polycarbonate submitted to shear loading. The numerical prediction is compared with the experimental results from Ghorbel et al. [60] and G'Sell et al. [62].The Figure 3a shows that the yield stress is slightly overestimated due to the difficulties of the cooperative model to fit both uniaxial and shear loadings. The shear behavior of polymers is governed by specific physics which should be taken into account here to obtain better prediction of the yield stress. From 40% strain, discrepancy between the numerical prediction and the experimental results appears. At 100% strain, a large overestimation of the numerical response in shear loading is observed (+120% at 100% strain) as mentioned by Ghorbel and collaborators [59, 60]. The Figure 3b shows the generalized model results. To fit the uniaxial compression while accounting for the true (3D) kinematics, a new set of parameters was identified (see Table 2). Using these data, the numerical prediction of the shear modelling was plotted on Figure 3b (red curve). For large strain, a slight underestimation occurs due to the change in stiffness in the experimental results around 70% strain (see [60, 62]). However, a very good agreement is observed between the numerical prediction and the experimental results of Ghorbel et al. [60] and G'Sell et al. [62] up to 70% strain after the parameters re-identification.

5. Conclusion

This paper presents the influence of the true kinematics of the deformable solid on the prediction of the three-dimensional mechanical modelling of amorphous polymers. This concept is not taken into account in the numerical mechanical model developed in the literature which leads to misjudgments of the stress tensor and rate of deformation tensor at large strain. To better understand its influence on the mechanical behavior of polymers, the true kinematics were implemented in the numerical model developed by Richeton et al. [26] through the logarithmic strain measure for elastic deformation, and the Green-Lagrange measure for viscoplastic deformation. It has been observed that the true kinematics influences the numerical prediction of the mechanical behavior of amorphous polymers submitted to both uniaxial and shear loadings. Because the material parameters are generally fitted on uniaxial tests (compression or tension), a new identification of the model parameters at large strain is needed to obtain a good agreement between experimental results and the generalized model predictions

**Funding**

The authors would like to acknowledge the French National Research Agency, under the name ANR-COSICO, for providing the funding for the post-doctoral fellowship of C. A. Bernard.

**Declaration of conflicting interests**

The authors declare that there is no conflict of interest.